# DIFFRACTIVE DEEP-INELASTIC SCATTERING °

R. Fiore[a†], L. L. Jenkovszky[b‡], F. Paccanoni[c*]

[a] *Dipartimento di Fisica, Università della Calabria,*
*Istituto Nazionale di Fisica Nucleare, Gruppo collegato di Cosenza*
*Arcavacata di Rende, I-87030 Cosenza, Italy*

[b] *Bogoliubov Institute for Theoretical Physics,*
*Academy of Sciences of the Ukrain*
*252143 Kiev, Ukrain*

[c] *Dipartimento di Fisica, Università di Padova,*
*Istituto Nazionale di Fisica Nucleare, Sezione di Padova*
*via F. Marzolo 8, I-35131 Padova, Italy*

**Abstract**

Diffractive deep inelastic events with a large rapidity gap are analyzed by using a Regge model for the pomeron flux and a gluonic content for the pomeron. Contrary to the expectations, the simplest assumption for the pomeron trajectory gives the best agreement with the data on the ratio of diffractive to the total number of events. In this case the main properties of the model are described by an analytic expression.

°*Work supported in part by the Ministero italiano dell'Università e della Ricerca Scientifica e Tecnologica and in part by the EEC Programme "Human Capital and Mobility", Network "Physics at High Energy Colliders", contract CHRX-CT93-0537 (DG 12 COMA)*

†*email address:* FIORE @CS.INFN.IT
‡*email address:* JENK @GLUK.APC.ORG
*\**email address:* PACCANONI @PADOVA.INFN.IT

## 1. Introduction

Recently events with a large rapidity gap have been revealed at HERA [1] that are not expected on the basis of standard deep-inelastic Montecarlo programs. In this paper we show that the experimental results are compatible with a theoretical model based on the diffractive nature of these events.

By definition, the differential cross section of a diffractive deep-inelastic process (Fig. 1) is

$$\frac{d\sigma(ep \to epX)}{d\xi dt dx dQ^2} = \frac{4\pi\alpha^2}{xQ^4}(1 - y + y^2/2) F_2^{diffr.}(x, Q^2; \xi, t) ,\qquad(1)$$

where $y$ is the pseudorapidity, $x = Q^2/2q\cdot p$ is the fraction of the proton momentum, $\xi = 1 - x_F$, $x_F = p'_z/p_z$ and $t$ is the squared proton momentum transfer, $t = (p-p')^2$. The $z$-axis points along the direction of the incoming proton. Note that $\xi \geq x$ and that, for a diffractive event $\mid t \mid \leq$ (few hundred MeV$^2$) and $\xi \ll 1$. We remind also that at HERA, $(k + p)^2 = \sqrt{s} = 296 GeV$ and the pseudorapidity of the smallest detector angle is $\eta = 4.3, \Theta = 1.5°$. The cut in $\eta$, $\eta_{max} < 1.5$, distinguishes events with a large rapidity gap and is equivalent to $\xi_{max} \leq 0.06$. Due to acceptance cuts, the maximum value of $\xi$ is $\xi_0 = 2.0 \times 10^{-2}$ for ZEUS [1].

Factorization

$$F_2^{diffr.}(x, Q^2; \xi, t) \to F_{P/p}(\xi, t) G_{q/P}(x/\xi, Q^2) \qquad(2)$$

of the structure function, where $F_{P/p}$ is the so called pomeron flux (see below) and $G_{q/P}$ is the pomeron structure function, is an important assumption [2-5] adopted in most of the calculations of diffractive deep inelastic scattering. We believe that this assumption does not need modifications. The distribution of the invariant mass of the hadronic system produced is peaked at small mass values and the triple-pomeron vertex, that appears only in the triple-Regge limit, does not contribute in this case.

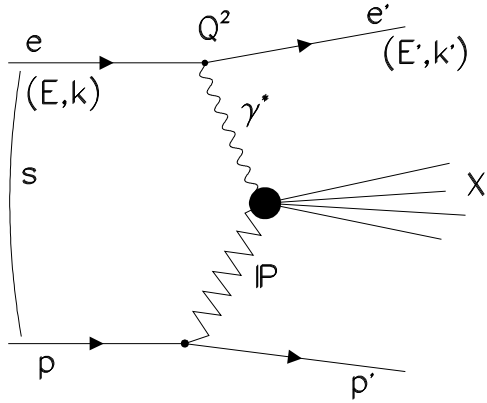

Figure 1: *Diagram for the diffractive deep inelastic scattering.*

## 2. The pomeron flux

The pomeron flux

$$F_{P/p} = \frac{1}{\sigma_{tot}(Pp)} \frac{d\sigma^{diff.}}{d\xi dt}$$

has been introduced [6] in the description of the diffractive dissociation.

We shall investigate the role of the $t$-dependence by using also non linear pomeron trajectories, fitted earlier [7] to elastic $pp$ and $\bar{p}p$ scattering. More important, we shall test also the role of the energy dependence in the pomeron flux. We remind that in earlier papers a constant value for the pomeron-proton total cross section was used (e.g., 2.3 mb in [8] or 1 mb in [6]). We shall include the energy dependence in the pomeron flux explicitly.

We start from the following simple form for the pomeron flux $F_{P/p}$, borrowed from our experience in describing high energy elastic hadron scattering

$$F_{P/p}(\xi, t) = [\exp(B\alpha(t))\xi^{-\alpha(t)}]^2 \xi ,\qquad(3)$$

where $\alpha(t)$ is the pomeron trajectory. In the case of a unit intercept linear trajectory $\alpha(t) = 1 + \alpha' t$ and for $B = 10$ [9], the expression (3) is numerically equivalent in the $t$-range under consideration to that for the pomeron flux used by Donnachie and Landshoff [10], even if the analytic form of the residue functions is different. By using a non linear pomeron trajectory, we shall study the effect of the non-exponential behaviour in $t$ as well.

The variation with $\xi$ of the pomeron flux may be studied by letting $\alpha(0)$ to be slightly beyond 1 ("supercritical" pomeron) or by using a dipole pomeron instead of Eq.(3), resulting in a logarithmic rise of the pomeron flux in $\xi$.

Practically, the pomeron flux cannot be completely isolated from non leading contributions that modify Eq.(3) as

$$F(\xi, t) = F_{P/p}(\xi, t) + a_R F_{R/p}(\xi, t) ,\qquad(4)$$

where

$$F_{R/p}(\xi, t) = [e^{(B_R \alpha_R(t))} \xi^{-\alpha_R(t)}]^2 \xi ,\qquad(5)$$

$\alpha_R(t)$ is an effective Regge trajectory and $a_R$ is the relative contribution of the "Reggeons". In this paper we will not consider the contribution from "secondary Reggeons". As will be shown elsewhere their presence does not change sensibly our conclusions.

### 3. Evolution of the pomeron structure function

To calculate the effect of the $Q^2$-evolution of the pomeron structure function we use a very simple input distribution corresponding to a pomeron made of a small number of ocean gluons only. In this case, and for non-large values of $Q^2$ ($Q^2 = 5 \div 10 GeV^2$), we assume a gluon distribution of the form [8]

$$zG(z, Q_0^2) = 2(1-z) , \qquad (6)$$

satisfying the momentum sum rule

$$\int_0^1 dz \; zG(z, Q_0^2) = 1 . \qquad (7)$$

We notice that Eq.(6) does not contrast with the experimental results [11] since, till now, the data cannot discriminate between the two choices $1-z$ an $z(1-z)$. Radiative corrections to the pomeron gluonic ladder suggest that the form (6) could be preferable but, at this stage, it can be considered as an ansatz.

The solution of the evolution equation in a wide range of $z$ necessitates cumbersome calculations and assumptions that make it difficult to get further transparent results for the diffractive DIS. In particular, it becomes problematic to take into account gluon recombination effects for the initial distribution (6) [3]. Expression (6) is fairly simple to enable an estimate of the evolution that will reproduce the main features of the gluon distribution in the kinematical region of interest, where $z = \frac{x}{\xi}$ assumes not too small values ($z \geq 0.01$), being $x \sim 0.001$ and $\xi_{max} \sim 0.06$.

Thus, we start from the distribution

$$zG(z, Q^2) = a(Q^2)(1-z)^{b(Q^2)} , \qquad (8)$$

in the kinematical region under consideration, with $a(Q_0^2) = 2$ and $b(Q_0^2) = 1$.

The expression for $b(Q^2)$ can be easily obtained near $z = 1$, when quarks are neglected, by going to the moments. For $z \approx 1$ the threshold behaviour of $zG(z, Q^2)$ is implemented by the large $n$ dependence of the anomalous dimension at the one-loop approximation [12]. We get

$$b(Q^2) = 1 + \frac{12}{11 - \frac{2N_f}{3}} s , \qquad (9)$$

with

$$s = ln(\frac{\ln(Q^2/\Lambda^2)}{\ln(Q_0^2/\Lambda^2)}) ,$$

$$Q_0^2 = 5 GeV^2, \quad \Lambda = 0.2 GeV .$$

Increasing $Q^2$ from $Q_0^2$ will diminish the value of the momentum integral (7). If we write $a(Q^2) = 2\exp(\alpha s)$ the bound

$$e^{\alpha s} < 1 + \frac{6}{11 - \frac{2N_f}{3}} s \qquad (10)$$

must be satisfied.

According to Ref. [13] the momentum integral is reduced by 10% at $Q^2 = 50\, GeV^2$ for the distribution we are considering. This condition determines $\alpha$ and the approximation we will use in this paper. With three flavors, $N_f = 3$, we obtain

$$b(Q^2) \approx 1 + \frac{4s}{3}, \ a(Q^2) \approx 2e^{s/3} \qquad (11)$$

and the resulting evolution for $z \geq 0.01$ is shown in Fig. 2 Let us now calculate the probability that a pomeron contains a parton with a momentum fraction $z$ from the evolution equation

$$\frac{\partial q(z, Q^2)}{\partial \ln(Q^2/\Lambda^2)} = \frac{\alpha_s}{2\pi} \int_z^1 \frac{dy}{y^2} yG(y, Q^2) P_{qg}(z/y) , \qquad (12)$$

with

$$P_{qg}(z/y) = 1/2 - z/y + z^2/y^2$$

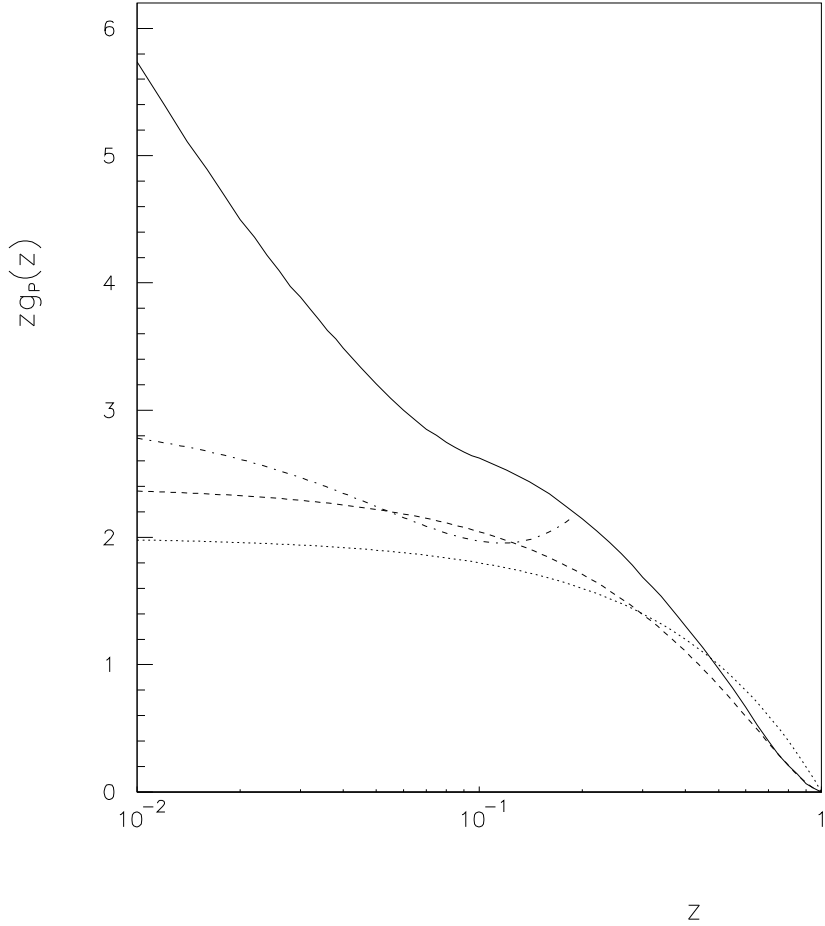

Figure 2: $z$-dependence of the gluon distributions: the initial distribution at $Q^2 = 5\ GeV^2$ (dotted line) is evolved to $Q^2 = 50\ GeV^2$ (dashed line); a sketch of the Ingelman-Prytz solution [2] is also shown (dotted-dashed line) together with the standard DGLAP evolution (full line) evaluated in Ref. [2].

and
$$yG(y, Q^2) = a(Q^2)(1-y)^{b(Q^2)},$$

where $a(Q^2)$ and $b(Q^2)$ are known.

Within the approximation suggested by Ingelman and Prytz [2], we obtain from Eq.(12) for $N_f = 3$ (see Appendix) the crude estimate

$$G_{q/P}(z, Q^2) =$$
$$\frac{8}{27}(1-z)^{b+1}a(Q^2)\left[1/3 + \frac{2b-1}{6}z - \frac{(2-b)(b-1)}{6}z^2\right.$$
$$\left. - \frac{zb}{2(b+1)}\left(1 + (b-1)z - \frac{(b-1)(2-b)}{3}z^2\right){}_2F_1(1, b+1; b+2; 1-z)\right]. \quad (13)$$

Other more rigorous approaches exist, see for example Ref. [3], but at the price of introducing one new parameter.

## 4. A simple model

Before starting with the numerical calculation, it may be useful to understand better the problem in a simplified model. Eq.(13) shows that a simple input for the pomeron flux will allow for an analytical solution. In Eq.(3)) we introduce a linear trajectory with $\alpha' = 0.25\ GeV^{-2}$ and write

$$F_{P/p}(\xi, t) = C(e^{B(1+\alpha't)}\xi^{-\alpha(t)})^2 \xi =$$
$$Ce^{2B}e^{2B\alpha't}\xi^{-1-2\alpha't}, \quad (14)$$

normalized according to Ref. [6] at $t = 0$ to

$$F_{P/p}(\xi, 0) = 3.536/\xi \rightarrow Ce^{2B} = c = 3.536 GeV^{-2}, \quad (15)$$

whence

$$F_{P/p}(\xi, t) = \frac{c}{\xi}e^{2\alpha'(B-\ln\xi)t}.$$

Integration over $t$ gives

$$\int_{-t_0}^{0} F_{P/p}(\xi, t)dt = \frac{c}{2\alpha'\xi(B - \ln \xi)}(1 - e^{-2\alpha'(B-\ln \xi)t_0}) \ . \tag{16}$$

Now, since the numerical value of the exponent

$$2\alpha'(B - \ln \xi)t_0$$

is typically 0.4 for $t_0 \sim 0.05 GeV^2, \alpha' = 0.25 GeV^{-2}$ and $\xi = 0.0009$ (note that the integration over $\xi$ extends from $x$ to $\xi_0 \sim 0.02$ according to the acceptance cut in ZEUS [1]), we can approximate Eq.(16) as

$$\int_{-t_0}^{0} F_{P/p}(\xi, t)dt \simeq \frac{ct_0}{\xi} \ . \tag{17}$$

To evaluate the relative contribution of diffractive deep inelastic events at HERA, we first have to calculate the quantity

$$L(x, Q^2) = \int_{-t_0}^{0} dt \int_{x}^{\xi_o} d\xi F_2^{diffr.}(x, Q^2; \xi, t) = ct_0 \int_{x}^{\xi_0} \frac{d\xi}{\xi} G_{q/P}(x/\xi, Q^2) \simeq$$

$$\frac{8}{27} ct_0 a(Q^2) \int_{x/\xi_0}^{1} dz(1-z)^{b+1} \left( \frac{1}{3z} + \frac{2b-1}{6} - \frac{(2-b)(b-1)}{6} z \right.$$

$$\left. -\frac{b}{2(b+1)}(1 + (b-1)z - \frac{(b-1)(2-b)}{3} z^2)\, _2F_1(1, b+1; b+2; 1-z) \right) \ . \tag{18}$$

Performing the integration in the right hand side of Eq.(18) (see Appendix for this purpose) leads to the result

$$L(x, Q^2) \simeq \frac{8}{81} ct_0 a(Q^2) \left(1 - x/\xi_0\right)^{b(Q^2)+2} \left[ \ln(\xi_0/x) \right.$$

$$+ \frac{(b(Q^2) - 1/2)(b(Q^2)^2 - 1/2)}{3(b(Q^2) + 2)} - \frac{(b(Q^2) - 1)(b(Q^2) - 2)}{6}(1 - x/\xi_0)$$

$$\left. -\frac{b(Q^2)}{2(1 - x/\xi_0)} \left( 3/2 - \frac{3x}{\xi_0} \ln(\frac{\xi_0}{x}) + \frac{(b(Q^2) - 1)(b(Q^2) - 2)}{3(b(Q^2) + 1)} \right) \right] \ , \tag{19}$$

where $c = 3.536 GeV^{-2}, \xi_0 = 0.02, a(Q^2) = 2e^{s/3}, b(Q^2) = 1 + 4s/3$. By using the parametrization (10) of Ref. [14] for $F_2(x, Q^2)$ we are now able to calculate the ratio

$$r = \frac{L(x, Q^2)}{F_2(x, Q^2)} \ . \tag{20}$$

The results of our calculations for various values of $x$ and for $Q^2$ in the interval between 10 and 100 $GeV^2$ and $t_0 = 0.05$ $GeV^2$ are shown with full lines in Fig. 3, where also the ZEUS data [1] are shown. Let us notice that we can arrive to the same results normalizing Eq.(14) according to Berger et al. [8] and choosing $t_0 \sim 0.1$. The independence of our finding from the parametrization chosen is exhibited in Fig. 4 where, by using Eq.(19), we fit preliminary data for the $F_2^{diffr.}(x, Q^2)$ [15].

## 5. Alternative models

A numerical analysis has been performed with the non linear trajectory

$$\alpha(t) = \alpha_0 + \alpha' t - \gamma(\sqrt{t_s - t} - \sqrt{t_s}) \tag{21}$$

where

$$\alpha' = 0.25 \ GeV^{-2} \ ,$$

$$\gamma = 0.03 \ GeV^{-1} \ ,$$

$$t_s = 4m_\pi^2 \simeq 0.078 \ GeV^2 \ .$$

When $t << t_s$ the trajectory (21) reduces to a linear one but, beyond the threshold $t_s$, it describes the transition to the Orear regime of the amplitude [16]. The virtue of this parametrization is that it accounts also for the small-$t$ structure observed in the diffraction cone in high energy hadron scattering (the so-called "break"). In ref. [16] more details and earlier references are given.

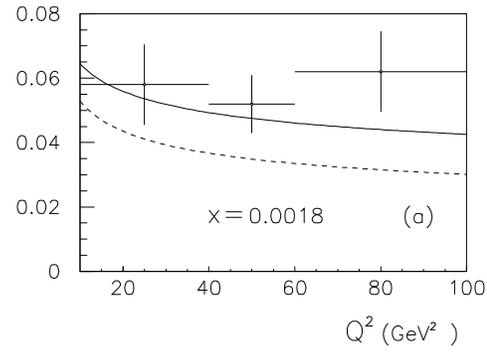
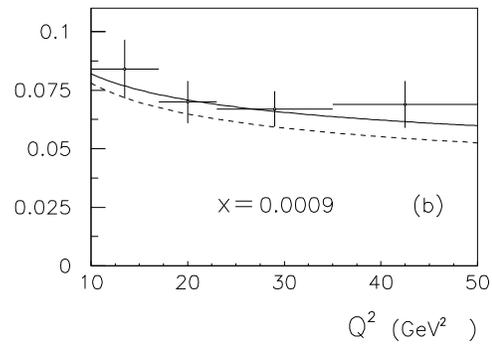
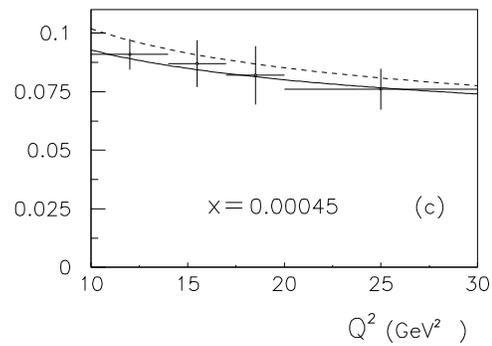

Figure 3: *The ratio r versus $Q^2$ for $x = 0.0018$ (a), 0.0009 (b), 0.0045 (c) and for $t_0 = 0.05$ and $\xi_0 = 0.02$, compared with the ZEUS data [1]: linear model (full lines) and supercritical pomeron with non linear trajectory (dashed lines).*

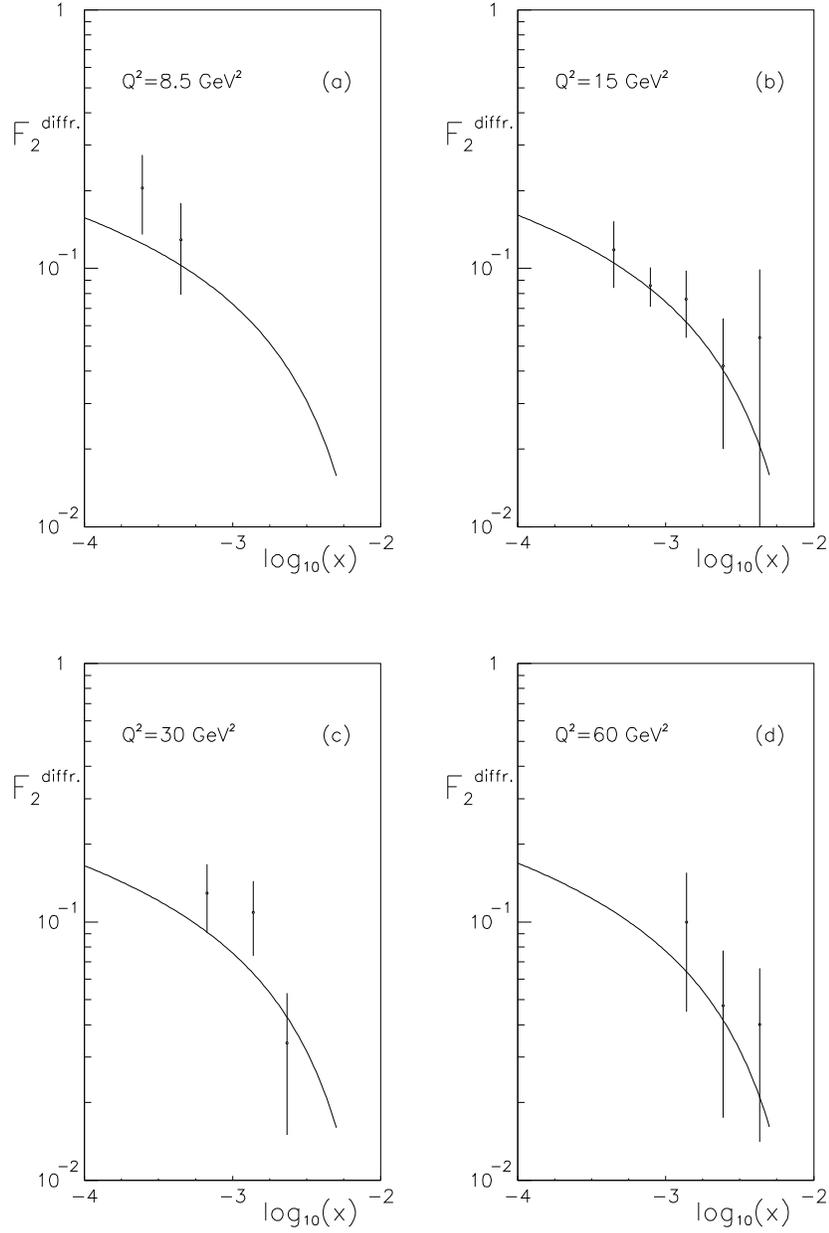

Figure 4: $F_2^{diffr.}(x, Q^2)$ versus $\log_{10}(x)$ for $Q^2 = 8.5\ GeV^2$ (a), $Q^2 = 15\ GeV^2$ (b), $Q^2 = 30\ GeV^2$ (c) and $Q^2 = 60\ GeV^2$ (d), as given in Eq.( 19) of the text, compared with the data from Ref. [15]

In a first instance the value of $\alpha_0$ has been set equal to one. The calculation provides also a test for the simple model of Sec. 4. The difference in the results as compared to the case of linear trajectories was expected to be small, since the $|t|$ integration range was restricted to a limited region near the forward direction and also because the non linear effects in the trajectory in this region are small anyway (numerically, $\gamma \ll \alpha'$). Surprisingly, the calculated effect in $r$ is considerably large and, in general, it tends to deteriorate the agreement with the available data.

The same is true for the pomeron intercept. Since the integration range in the energy variable (pomeron flux) is limited, one would expect that the choice of the pomeron singularity (i.e. the value of the pomeron intercept) has little effect on the final results. In fact, this is not the case: the introduction of a new parameter, $\delta = \alpha(0) - 1$ (in our calculations we put $\delta = 0.1$) again deteriorates the results. Finally the results are not better in the case of a "mixed" option, i.e. when both the trajectory is non linear and the pomeron intercept is "supercritical". As an example, we show in Fig. 3 (dashed lines) the result of the computer calculation for the ratio $r$ with a supercritical pomeron, $\delta = 0.1$ and the non-linear trajectory (21).

It should be stressed that by working with non linear trajectories and a supercritical pomeron, we have not introduced free parameters - their values were taken from earlier fits of elastic hadron scattering [7] and were not varied here. Since, however, those fits and the relevant parameters themselves may vary, this point needs further clarification. Here we can only assert that good agreement has been achieved in the "minimal" model of a linear trajectory and simple pomeron pole with a unit intercept (constant asymptotically total cross sections) and that the final results are sensitive even to small variations of the parametrization of the "pomeron flux".

## 6. Conclusions

The aim of the present paper was to study the pomeron internal structure by using diffractive deep inelastic scattering date from HERA [1]. We have constructed a model for the diffractive deep inelastic scattering based on the formalism developed earlier [2] as well on our present knowledge about the elastic scattering amplitude (pomeron exchange) and the assumed pomeron structure (distribution functions). By comparing the calculated $Q^2$ dependence (at various fixed values of $x$) of the relative contribution of the diffractive DIS events, Eqs.(17) and (18) to the HERA data, the role of the various components of the input, namely the pomeron structure function, the form of the pomeron trajectory and its intercept, as well as the effect of various approximations used in the calculations were tested.

We have tried different inputs as regards the form and the intercept of the pomeron trajectory. A non linear trajectory has been also considered that, for hadronic reactions, describes the small-$t$ "break" in the differential cross-section. Besides the unit intercept, the case of a supercritical one, with $\alpha(0) = 1.1$ has been analyzed too.

While the results of the calculations for different inputs, including also mixed options, sensibly differ, the main conclusions are quite general. We found noticeable dependence on the pomeron intercept, which means that, in spite of the limited interval for the energy (pomeron flux) variation, its rate has important consequences for the ratio $r$.

The introduction of non-linear trajectories does not improve the resulting behaviour of the ratio $r$. A better choice of the parameters may improve the agreement with the data, but anyway this is only a small effect as compared to the available freedom in choosing the pomeron structure function.

While the variation of the pomeron flux is meant mainly "to fix normalization", the objective of the present paper - like in similar investigations by other authors [2, 3, 6, 8, 10] - was the right choice of the pomeron structure function, i.e. the deduction of the pomeron internal structure from the data. We have assumed that the pomeron is made of a small number of gluons distributed according to Eq.(6), with the parameters $a(Q^2)$ and $b(Q^2)$, calculated from the evolution equation, given by

$$b(Q^2) \simeq 1 + 4\frac{s}{3}, \qquad a(Q^2) \simeq 2e^{\frac{s}{3}}$$

and normalized according to $a(Q_0^2) = 2$, $b(Q_0^2) = 1$. Note that the above distribution is close to the one calculated by G.Ingelman and K.Prytz [2] taking into account gluon recombination. Any modification of the pomeron distribution by using e.g.

$$zG(z, Q_0^2) = 6(1-z)^5$$

(corresponding to the presence of many gluons inside the pomeron) instead of Eq.(6) and its QCD evolution, sensibly affects the resulting behaviour of the ratio $r$. The large errors in the experimental diffractive structure functions do not allow, however, to exclude this possibility. The agreement with the data of Ref.[1] of the results of our calculations is quite satisfactory since, for a given choice of the pomeron structure function, practically no parameters were introduced (those used in our calculations were fixed elsewhere - for example the pomeron trajectory was adjusted earlier to fit hadronic cross sections). This agreement can be made even better by letting some of the free parameters to vary and to be fitted to the diffractive deep inelastic data, which however are still quite preliminary and subject to further improvement.

To summarize, we consider our model as a laboratory for studying the pomeron structure function by comparing the calculated results with the experimental data. In this paper we made the first step by proving the stability of our model against the variations of its component and tentatively using a rather simple gluon distribution function for the pomeron internal structure.

Acknowledgement: We thank M. Arneodo, A. Garfagnini, L. Iannotti and A. Solano for useful discussions on the subject of this work. One of us (L. L. J.) is grateful to the Dipartimento di Fisica dell'Università della Calabria and to the Istituto Nazionale di Fisica Nucleare - Gruppo collegato di Cosenza for their warm hospitality and financial support.

## 6. Appendix

Let us start from Eq.(11). By using the Ingelman-Prytz approximation [2], for $N_f = 3$ we get

$$q(z, Q^2) = \frac{\alpha_s}{2\pi} \int_z^1 \frac{dy}{y^2} a(Q^2)(1-y)^{b(Q^2)}(1/2 - z/y + z^2/y^2) \ln(Q^2/\Lambda^2) =$$

$$2/9 a(Q^2) \int_z^1 dy/y^2 (1-y)^{b(Q^2)}(1/2 - z/y + z^2/y^2) \ .$$

The integration can be done explicitly yielding

$$I = \int_z^1 dy/y^2 (1-y)^{b(Q^2)}(1/2 - z/y + z^2/y^2) = 1/2 I_2 - z I_3 + z^2 I_4 \ ,$$

where
$$I_n = \int_0^{1-z} dt \frac{t^b}{(1-t)^n} = \frac{(1-z)^{b+1}}{b+1} {}_2F_1(n, b+1; b+2; 1-z) .$$

By applying to ${}_2F_1(n, b+1; b+2; 1-z)$ the recurrence formula
$$F(n) = \frac{1}{n-1}((b+1)z^{-n+1} + (n-b-2)F(n-1)) ,$$

we get
$$I = \left( \frac{(b+1)}{3z} + \frac{(b+1)(2b-1)}{6} - \frac{(2-b)(b^2-1)z}{6} \right.$$
$$\left. -1/2(b + b(b-1)z - b(b-1)(2-b)z^2/3) {}_2F_1(1, b+1; b+2; 1-z) \right) \frac{(1-z)^{b+1}}{b+1} .$$

The integration in $z$ may be easily done by using the form
$${}_2F_1(1, b+1; b+2; 1-z) = (b+1) \int_0^1 dt \frac{t^b}{1-t(1-z)} .$$

We can approximate the result with the limit
$$\lim_{z \to 1} \Phi(z, 1, v) / (-\ln(1-z)) = 1 ,$$

where
$$\Phi(z, 1, v) = \sum_{n=0}^{\infty} (v+n)^{-1} z^n .$$

To get
$$G_{q/P}(z, Q^2) = \sum_f e_f^2 [zq_f(z, Q^2) + z\bar{q}_f(z, Q^2)]$$

it is sufficient to multiply $zq(z, Q^2)$ by $4/3, (N_f = 3)$.